\documentclass[prd,preprint, nofootinbib,amsmath,amssymb, aps, floatfix]{revtex4-1}
\pdfoutput=1
\usepackage{dcolumn}
\usepackage{bm}
\usepackage{amssymb,amsmath,graphicx}
\usepackage[linktoc=page]{hyperref}
\usepackage{tikz}
\usepackage{fancyhdr}
\usetikzlibrary{decorations.pathmorphing}

\newcommand{\T}{\mathfrak{t}}
\newcommand{\A}{\mathcal{A}}
\newcommand{\C}{\mathcal{C}}
\newcommand{\V}{\mathcal{V}}
\newcommand{\x}{\mathbf{x}}

\begin{document}
\title{Evolution of Complexity Following a Global Quench }
\author{Mudassir Moosa}%
\email{mudassir.moosa@berkeley.edu}
\affiliation{ Center for Theoretical Physics and Department of Physics\\
University of California, Berkeley, CA 94720, USA 
}%
\affiliation{Lawrence Berkeley National Laboratory, Berkeley, CA 94720, USA}

\begin{abstract}

The rate of complexification of a quantum state is conjectured to be bounded from above by the average energy of the state. A different conjecture relates the complexity of a holographic CFT state to the on-shell gravitational action of a certain bulk region. We use `complexity equals action' conjecture to study the time evolution of the complexity of the CFT state after a global quench. We find that the rate of growth of complexity is not only consistent with the conjectured bound, but it also saturates the bound soon after the system has achieved local equilibrium. 

\end{abstract}
\maketitle
\tableofcontents

\section{Introduction}
\label{intro}


The quantum complexity of a state is defined as the minimum number of gates required to prepare this state starting from some reference state \cite{susskind-comp,Aaronson:2016vto}. The complexity was introduced in the context of AdS-CFT correspondence to describe the growth of the interior of a two-sided black hole from the CFT perspective. More precisely, it was proposed that the growth of the interior of the black hole is dual to the growth of the complexity of the dual CFT state \cite{susskind-comp}.

These ideas were made more precise in \cite{ca-1,ca-2}, where it was conjectured that the complexity $\C(\T)$ of the boundary state at time $\T$ is proportional to the value of the on-shell gravitational action $\A(\T)$ of a certain bulk region. This bulk region is the domain of dependence of a Cauchy slice anchored on the boundary at time $\T$ (see Fig.~(\ref{fig-1}) for an example). This conjecture is known as \textit{complexity equals action} (CA) conjecture, and the bulk region is called the Wheeler-deWitt (WdW) patch. The CA conjecture asserts that
\begin{equation}
\C(\T) = \frac{\A(\T)}{\pi} \, . \label{CA-conj}
\end{equation}

A universal bound, known as Lloyd's bound \cite{2000Natur.406.1047L}, conjectures that the energy in the system bounds the rate of computation by the system. Inspired by this, \cite{ca-1,ca-2} conjectured an analogous bound on the rate of the growth of the complexity. The bound states
\begin{equation}
\frac{d}{d\T} \C(\T) \le \, \frac{2}{\pi} \, E \, , \label{eq-Lloyd-bound}
\end{equation}
where $E$ is the average energy of the state at time $\T$. The original works \cite{ca-1,ca-2} performed various calculations to test this conjecture. {These works} used CA conjecture, Eq.~\eqref{CA-conj}, to find the rate of complexification at late times for isolated two-sided black hole and for two-sided black hole perturbed by shock waves \cite{shock-waves}. Since then, the CA conjecture has been studied in detail. It was shown in \cite{Yang:2016awy} that the rate of complexification for two-sided black hole satisfies the bound of Eq.~\eqref{eq-Lloyd-bound} at late times if the matter fields only appear outside the killing horizon. The bound is also saturated for BTZ black holes in minimally massive gravity model \cite{Qaemmaqami:2017lzs}.  The complexity of entangling two CFTs was computed in \cite{comp-formation} by calculating the  difference between the complexity of a thermo-field double state and the complexity of a product state of vacua of two CFTs. Recently, the late time analysis of \cite{ca-1,ca-2} was extended in \cite{Carmi:2017jqz,Kim:2017qrq}, where the full time dependence of the rate of complexification for two-sided black hole was calculated. It was shown in these papers that the conjectured bound of Eq.~\eqref{eq-Lloyd-bound} is violated at earlier times. Similar violations were also observed for the dual of the non-commutative gauge theories in \cite{Couch:2017yil}. A generalization of Eq.~\eqref{CA-conj} for the reduced state of some subsystem of the CFT was proposed in \cite{comp-subregions}.  The CA conjecture has {also sparked} interests to study quantum complexity for quantum field theories without making connection to holography \cite{Hashimoto:2017fga,myers,heller,Yang:2017nfn}.  

Even though the CA conjecture, Eq.~\eqref{CA-conj}, and the bound of the rate of complexification, Eq.~\eqref{eq-Lloyd-bound}, are not consistent in general, our goal in this paper is to check their consistency for quenched quantum system. The study of global quenches has gained considerable attention in recent years as a tool to understand how a closed quantum system equilibrates. The idea behind quantum quenches is to excite the system out of equilibrium and let it evolve under unitary time evolution. The global quenches have been extensively studied using both field-theoretic methods and holography \cite{gq-1,gq-2,gq-3,Albash:2010mv,gq-4,gq-5,gq-6,gq-7,gq-8,gq-9,gq-10,gq-11,gq-12,gq-13,gq-14,gq-15,gq-16,gq-17,gq-18,gq-19,gq-20,gq-21,gq-22}. Previous works on this topic focus on studying the time evolution of the correlation functions, Wilson loops, and entanglement entropy of subregions. Since the complexity does not stop growing even after system has achieved equilibrium, we do not expect to get new insights about the system's approach to equilibrium. Nevertheless, we expect quantum quenches to provide a non-trivial test of the bound in Eq.~\eqref{eq-Lloyd-bound}. 

Our approach in this paper is to use AdS-CFT correspondence to study the quantum quench of the CFT. This allows us to use the CA conjecture, Eq.~\eqref{CA-conj}, to calculate the complexity of the boundary state. Recall that the time evolution of the quenched CFT state and its approach to thermalization can be described by a black hole formed by a collapse in the bulk \cite{gq-3,gq-4,gq-6,gq-7,gq-8,gq-10,gq-11,gq-13,gq-14}. The Penrose diagram for this spacetime is shown in Fig.~(\ref{fig-1}), where the WdW patch corresponding to boundary state at time $\T$ is shown as the shaded region. We describe in detail the quench protocol and the bulk geometry in Sec.~(\ref{setup}).

We carry out the main calculations of this paper in Sec.~(\ref{calculations}). { Here, we summarize the results of our calculation.} For two-dimensional CFTs, we find the full time-dependence of the complexity of the quenched state (see Eq.~\eqref{comp-d2}). For CFTs in higher spacetime dimensions, we derive an analytical expression for the rate of complexification as a function of time. Our expression for the rate of complexification is in terms of the position of the codimension-$2$ plane where the null boundary of the WdW patch intersects the infalling shell.  This plane is labeled $P$ in Fig.~(\ref{fig-1}) and Fig.~(\ref{fig-2}). We find that the fact that the plane $P$ never crosses the event horizon is equivalent to the bound of Eq.~\eqref{eq-Lloyd-bound}. This provides a non-trivial consistency check of the CA conjecture, Eq.~\eqref{CA-conj}, and the conjectured bound, Eq.~\eqref{eq-Lloyd-bound}, for quenched quantum system. We further find, using numerical analysis, that the growth of complexity of the quenched state saturates this bound 
when the system has approached local equilibrium.

We end with {a summary} and some possible extensions of our work in Sec.~(\ref{disc}).


\section{Setup} \label{setup}

We consider a CFT with a holographic dual in a $d$-dimensional flat spacetime. We start with the vacuum state of this CFT and at time $\T=0$ globally \textit{quench} the system  by suddenly injecting energy with finite  density everywhere in the system. We can achieve this by turning on a global source for a small amount of time. We assume that the quench is homogeneous and isotropic to preserve the symmetries of the system. The state of the system after the quench is no longer an eigenstate of the Hamiltonian of our CFT. Hence, it evolves with time until it equilibrates. 

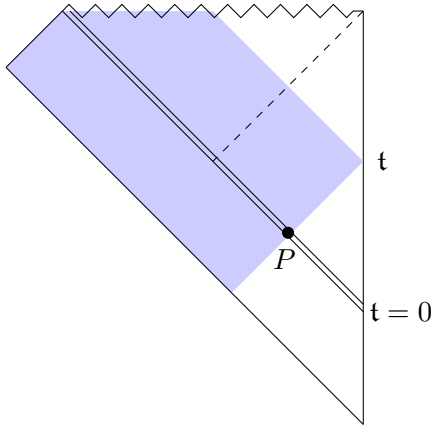
\begin{figure}[h]
		\begin{tikzpicture}
		\draw [black](0,-1.5) --(0,4);
		\fill [blue!20!white](0,2) --(-2,4) --(-4,4) --(-4.75,3.25)--(-1.75,0.25)--(0,2) ;
		\draw [black](0,-1.5) --(-4.75,+3.25);
		\draw [black](-4,4) --(-4.75,3.25);
		\draw [black](0,0) --(-4,4);
		\draw [black](0,0.1) --(-3.9,4);
		\draw [decorate,decoration=zigzag](-4,4) --(0,4);
		\draw [dashed](0,4) --(-2,2);
		\node at (.5,0) {$\T=0$};
		\node at (.25,2) {$\T$};
		\node at (-1.05,0.7) { {\small $P$}};
		\fill [black](-1.0,1.05) circle (.08cm);
		\end{tikzpicture}
		\caption{The Penrose diagram of the time dependent geometry as a result of a collapse of a null shell (shown as a double line). The dashed line is the event horizon, and the shaded region denotes the WdW patch corresponding to boundary time $\T$. The intersection of the past null boundary of the WdW patch and the collapsing null shell is denoted by a black dot and is labeled by $P$.}
		\label{fig-1}
\end{figure}

The global quench and the approach to thermalization of the quenched state have a simpler description in the bulk side of the AdS-CFT correspondence. The quench in the CFT corresponds to introducing a planar `shell' of matter near the boundary of the AdS spacetime. This shell then collapses to form a black hole in the bulk, which represents the equilibrium state in the boundary CFT. We follow \cite{gq-3,gq-4,gq-6,gq-7,gq-8,gq-10,gq-11,gq-13,gq-14} and model the global quench by a $(d+1)$-dimensional planar AdS-Vaidya metric \footnote{We set $\ell_{AdS}=1$.},
\begin{equation}
ds^{2} = \frac{1}{z^{2}} \, \Big( -f(v,z) \, dv^{2} - 2 dv dz + d\x^{2}_{d-1}\Big) \, , \label{eq-vaidya}
\end{equation}
where 
\begin{align}
f(v,z) = 
\begin{cases}
1 &\text{for $v <0$} \, ,\\[0ex]
1 - \dfrac{z^{d}}{z_{h}^{d}} &\text{for $v>0$} \, , \label{eq-fvz}
\end{cases}
\end{align}
and $z=z_{h}$ is the position of the event horizon. This metric describes the AdS-Schwarzschild black hole for $v>0$ and the vacuum AdS for $v<0$. The Penrose diagram of this spacetime is shown in Fig.~(\ref{fig-1}).

The parameter $z_{h}$ can be interpreted as the \textit{local equilibrium scale} after which time the system achieves local equilibrium \cite{gq-10}. The temperature and energy of the equilibrium state in the boundary are same as those of the black hole formed in the bulk and are given in terms of $z_{h}$ as
\begin{equation}
T_{d} = \, \frac{d}{4\pi\, z_{h}} \, , \,\,\,\,\,\,\,\,\,\,\,\,\,\,\,\,\,\,\,\,\,\,\,\,\, E_{d} = \, \frac{d-1}{16\pi G} \, \frac{L_{d-1}}{z_{h}^{d}} \, , \label{eq-E}
\end{equation}
where we have defined 
\begin{equation}
L_{d-1} \equiv \int d^{d-1}x \, , \label{eq-size}
\end{equation}
as the volume of our boundary system.

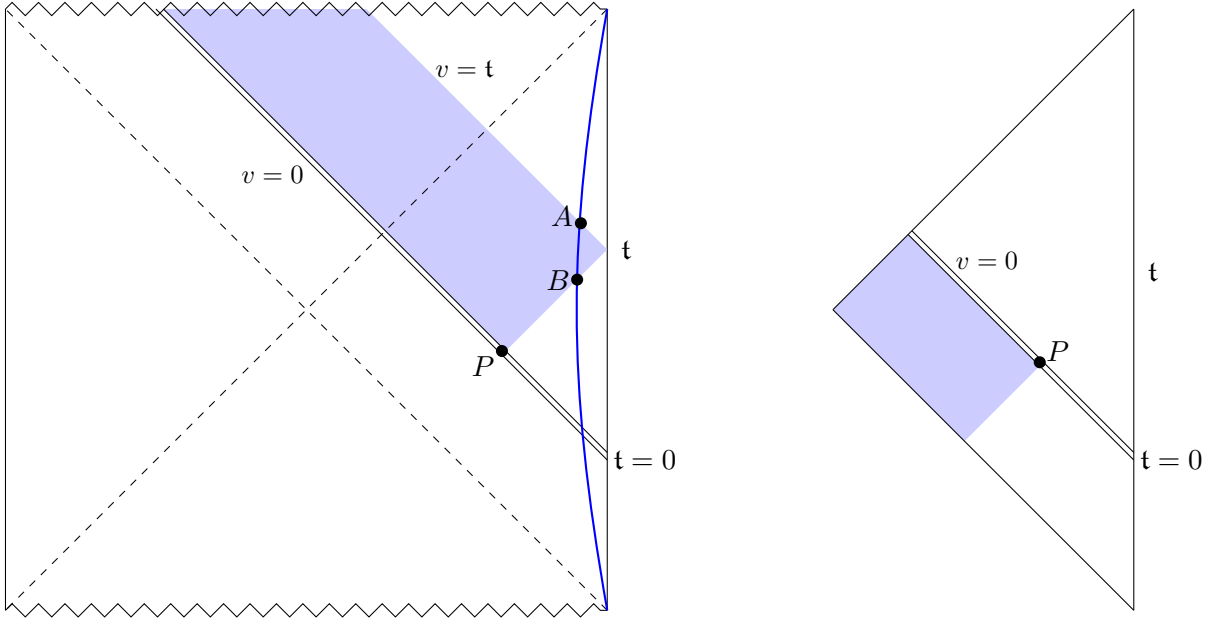
\begin{figure*}[h]
	\begin{tikzpicture}
	\draw [white](0,4.05) --(-8,4.05);
	\draw [black](0,-4.0) --(0,4);
	\draw [black](-8,-4.0) --(-8,4);
	\fill [blue!20!white](0,.8) --(-3.2,4) --(-5.9,4) --(-1.35,-0.55)--(0,.8) ;
	\draw [black](0,-2.0) --(-6.0,4);
	\draw [black](0,-1.9) --(-5.9,4);
	\draw [decorate,decoration=zigzag](-8,4) --(0,4);
	\draw [decorate,decoration=zigzag](-8,-4) --(0,-4);
	\draw [dashed](0,4) --(-8,-4);
	\draw [dashed](0,-4) --(-8,4);
	\node at (.5,-2.0) {$\T=0$};
	\node at (.25,0.8) {$\T$};
	\node at (-1.65,-0.75) { $P$};
	\fill [black](-1.4,-.55) circle (.08cm);
	\node at (-4.45,1.8) {{\footnotesize $v=0$}};
	\node at (-1.9,3.2) {{\footnotesize $v=\T$}};
	\draw [blue,thick](0,4) to [out=260,in=100] (0,-4);
	\node at (-0.6,1.25) {{\small $A$}};
	\fill [black](-0.35,1.150) circle (.08cm);
	\node at (-0.65,0.40) {{\small $B$}};
	\fill [black](-0.4,0.40) circle (.08cm);
	\draw [black](7,-4) --(7,4);
	\fill [blue!20!white](5.75,-.75) --(4.0,1.0) --(3,0) --(4.75,-1.75)--(5.75,-.75) ;
	\draw [black](7,-2.0) --(4.,1.);
	\draw [black](7,-1.9) --(4.05,1.05);
	\draw [black](7,4) --(3,0);
	\draw [black](7,-4) --(3,0);	
	\node at (7.5,-2.0) {$\T=0$};
	\node at (7.25,0.5) {$\T$};
	\node at (5.99,-0.55) { $P$};
	\fill [black](5.75,-0.70) circle (.08cm);
	\node at (5.05,0.65) {{\footnotesize $v=0$}};
	\end{tikzpicture}
	\caption{The Penrose diagrams for AdS-Schwarzschild black hole (left) and vacuum AdS (right). The shaded region on the left/right corresponds to the part of the WdW inside/outside the collapsing null shell in Fig.~(\ref{fig-1}). We have also included a cut-off surface at $z=\delta$ (shown as a blue line) in the Penrose diagram of AdS-Schwarzschild. The intersections of the WdW patch with the cut-off surface are labeled $A$ and $B$, whereas the intersection of the past null boundary of the WdW patch and the collapsing null shell is denoted by a black dot and is labeled $P$.}
	\label{fig-2}
\end{figure*}

Our goal is to to compute the complexity of the boundary state after the global quench as a function of time. According to the CA conjecture, Eq.~\eqref{CA-conj}, this complexity of the boundary state at time $\T$ is related to the action of the WdW patch corresponding to the boundary time $\T$. This patch is shown  as the shaded region in Fig.~(\ref{fig-1}). Note that the on-shell action of a null dust vanishes \cite{dust,null_dust}. Moreover, since the stress energy tensor only has a $T_{vv}$ component, the trace of the Einstein's equations implies that the Ricci scalar is non-singular at $v=0$. This allows us to use the additive property of the gravitational action \cite{Lehner:2016vdi}, and split our WdW patch into two parts. One that lies entirely inside the infalling null shell ($v > 0$) and the other that lies outside the null shell ($v < 0$). That is,
\begin{equation}
\A = \, \A_{(v>0)} + \A_{(v < 0)} \, , \label{eq-ac-total}
\end{equation}
where $\A_{(v>0)}$ and $\A_{(v<0)}$ are the gravitational actions of the shaded regions in the left and right columns of Fig.~(\ref{fig-2}) respectively. As a result, instead of working with time-dependent metric of Eq.~\eqref{eq-vaidya}, we can simply do our calculations in stationary AdS-Schwarzschild and vacuum AdS spacetimes. We carry out these calculations in the next section.

\section{Calculations} \label{calculations}

In this section, we perform the main calculations of this paper. In Sec.~(\ref{sec-bh}), we consider the part of the WdW patch that lies inside the infalling shell (left column of Fig.~(\ref{fig-2})) and calculate its action, $\A_{(v>0)}$. We then calculate the action $\A_{(v<0)}$ of the part of the WdW patch that lies outside the infalling shell (right column of Fig.~(\ref{fig-2})). We combine the results of these calculations in Sec.~(\ref{total-action}) and show that the rate of complexification, determined using the CA conjecture, Eq.~\eqref{CA-conj}, satisfies the bound of Eq.~\eqref{eq-Lloyd-bound}.

\subsection{Calculation of $\A_{(v>0)}$}\label{sec-bh}

The shaded region inside the collapsing shell has one spacelike, one timelike, and three null boundaries. The spacelike boundary is the singularity ($z=\infty$). The timelike boundary is the cut-off surface ($z = \delta$) that we have introduced to regulate the UV divergence in the gravitational action. Two of the null boundaries are $v = 0$ and $v = \T$, while the third null boundary is the plane joining points $B$ and $P$ in Fig.~(\ref{fig-2}). Let's assume that this null boundary is described by $z_{1}(v;\T)$. The function $z_{1}(v;\T)$ then satisfies the integral equation
\begin{equation}
\frac{1}{2}(\T - v) = \, \int_{0}^{z_{1}(v;\T)}  dz \, \frac{1}{f(z)} \, , \label{def-z1}
\end{equation}
where $f(z) = f(v>0,z)$ in Eq.~\eqref{eq-fvz}. The coordinates of the points $A$, $B$, and $P$ are
\begin{align}
A: \,\,\,\, v_{A} =& \, \T \, , \,\,\,\,\,\,\,\,\,\,\,\,\,\,\,\,\,\,\,\,\,\,\,\,\, z_{A} = \, \delta \, ,\label{eq-cor-A} \\
B: \,\,\,\, v_{B} =& \, \T - 2\delta \, , \,\,\,\,\,\,\,\,\,\,\,\, z_{B} = \, \delta \, , \label{eq-cor-B}\\
P: \,\,\,\, v_{A} =& \, 0 \, , \,\,\,\,\,\,\,\,\,\,\,\,\,\,\,\,\,\,\,\,\,\,\,\, z_{P}(\T) = \, z_{1}(0,\T) \, . \label{eq-cor-z}
\end{align}
In the case of $d=2$, the Eq.~\eqref{def-z1} can be exactly solved and the solution is given by
\begin{equation}
z_{1}(v;\T) = \, z_{h} \, \tanh \Bigg( \frac{\T - v}{2 z_{h}} \Bigg) \, , \,\,\,\,\,\,\,\,\,\,\,\,\,\,\,\,\,\,\,\,\,\,\,\,\,\,\,\,\,\,\,\, \text{for } d = 2 \, . \label{eq-z1-2}
\end{equation}
However, the closed form solution of Eq.~\eqref{def-z1} is not available for $d \ge 3$. 

We now {focus our attention to compute} the gravitational action $\A_{(v>0)}$ of the shaded region in the left column of Fig.~(\ref{fig-2}). The bulk contribution to the gravitational action is given by the Einstein-Hilbert term. Since the Ricci scalar of the AdS-Schwarzschild black hole is constant, the bulk contribution to $\A_{(v>0)}$ is proportional to the volume $\V_{(v>0)}$ of the shaded region in the left column of Fig.~(\ref{fig-2}). That is, 
\begin{equation}
\A_{(v>0)}^{\text{bulk}} = \, \frac{1}{16\pi G} \, \Big( R - 2 \Lambda \Big) \, \V_{(v>0)} \, , \label{eq-EH}
\end{equation}
where $R$ is the Ricci scalar and $\Lambda$ is the cosmological constant. The volume of the shaded region is given by
\begin{align}
\V_{(v>0)} =& \, \int d^{d-1}x \, \int_{0}^{\T-2\delta} dv \, \int_{z_{1}(v;\T)}^{\infty} dz \, \frac{1}{z^{d+1}} \, \nonumber\\ +&  \, \int d^{d-1}x \, \int_{\T-2\delta}^{\T} dv \, \int_{\delta}^{\infty} dz \, \frac{1}{z^{d+1}} \, , \nonumber\\
=& \, \frac{L_{d-1}}{d}\,\frac{2}{\delta^{d-1}}  + \, \frac{L_{d-1}}{d}\, \int_{0}^{\T-2\delta} dv \, \frac{1}{z^{d}_{1}(v;\T)} \, , \label{eq-vol1}
\end{align}
where we have used Eq.~\eqref{eq-size}. 

In the case of $d=2$, we can use Eq.~\eqref{eq-z1-2} to get
\begin{equation}
\V_{(v>0)}^{(d=2)} = \, \frac{2L_{1}}{\delta} + \frac{L_{1}}{2 z_{h}^{2}} \, \T - \frac{L_{1}}{z_{h} \tanh\Big(\frac{\T}{2z_{h}}\Big)} \, . \label{eq-vol1-d2}
\end{equation}
Since we do not have a closed form formula of $z_{1}(v;\T)$ for $d\ge 3$, we cannot perform the integrals in Eq.~\eqref{eq-vol1}. Regardless, we can still calculate the contribution of the Einstein-Hilbert term to the rate of complexification. To do this, we take the time derivative of the volume in Eq.~\eqref{eq-vol1}. We get
\begin{equation}
\frac{d}{d\T} \V_{(v>0)} = \, \frac{L_{d-1}}{d} \, \frac{1}{z_{P}^{d}(\T)} \, . \label{eq-der-1}
\end{equation}
In deriving this result, we have used the following property of the derivatives of $z_{1}(v;\T)$
\begin{equation}
\frac{d}{d\T} z_{1}(v;\T) = \, - \, \frac{d}{d v} z_{1}(v;\T) \, .
\end{equation}
Our approach in this and the next subsection is to write all the contributions to the gravitational action in terms of $z_{P}(\T)$. We then analyze these results in Sec.~(\ref{total-action}) where we numerically solve for $z_{P}(\T)$ for $d\ge 3$.

Recall that the cosmological constant is equal to $\Lambda = -\frac{1}{2}d(d-1)$. The vacuum Einstein's equations then imply $R = -d(d+1)$. We combine Eq.~\eqref{eq-EH} and Eq.~\eqref{eq-vol1-d2} to get
\begin{equation}
\A_{(v>0)}^{\text{bulk}} = \, \frac{L_{1}}{8\pi G} \Bigg( \frac{2}{z_{h} \tanh\Big(\frac{\T}{2z_{h}}\Big)} - \frac{1}{z_{h}^{2}} \, \T  - \frac{4}{\delta}  \Bigg) \, , \,\,\,\,\,\,\,\,\,\,\,\,\,\,\,\,\,\,\,\,\,\,\,\,\,\,\, \text{for } d = 2 \, ,
\end{equation}
and we combine Eq.~\eqref{eq-EH} and Eq.~\eqref{eq-der-1} to get
\begin{equation}
\frac{d}{d\T} \A^{\text{bulk}}_{(v>0)} = \, - \frac{L_{d-1}}{16\pi G} \, \frac{2}{z_{P}^{d}(\T)} \, .
\end{equation}

We now consider the boundary contributions to $\A_{(v>0)}$. There are two contributions to the action from a null boundary. The first contribution is proportional to the surface integral of the `surface gravity' $\kappa$ of the null generator $k^{a}$ of the null surface (see for \textit{e.g.} \cite{Lehner:2016vdi}). The second contribution is a counter term which ensures that the action is independent of a choice of the parameterization of the null surface \cite{Lehner:2016vdi,Reynolds:2016rvl}. The `surface gravity' $\kappa$ of the null generator $k^{a}$ is defined as
\begin{equation}
k^{a}\nabla_{a}k^{b} = \, \kappa \, k^{b} \, .
\end{equation}
Let's denote the generator of the null boundaries $v=0$ and $v=\T$ by $k_{\text{in}}^{a}$ and that of  the null boundary between points $B$ and $P$ in Fig.~(\ref{fig-2}) by $k_{\text{out}}^{a}$.
We choose the following normalization of the null generators
\begin{align}
k_{\text{in}}^{a} =& \, -z^{2} \, \big( \partial_{z}\big)^{a} \, ,\label{eq-kin}\\
k_{\text{out}}^{a} =& \, \frac{2 z^{2}}{f(z)} \, \big( \partial_{v}\big)^{a} - z^{2} \, \big( \partial_{z}\big)^{a} \, .\label{eq-kout}
\end{align}
The advantage of this particular normalization of the null generators is that the surface gravity of both of these null generators vanish. Therefore, we only have to focus on the counter terms. The contributions from the counter terms is \cite{Lehner:2016vdi}
\begin{align}
\A_{(v>0)}^{\text{null}} =& \, \frac{1}{8\pi G} \, \int d^{d-1}x \,  \int_{\delta}^{\infty} \, \frac{dz}{z^{2}} \, \sqrt{q} \, \theta_{\text{in}}\log\theta_{\text{in}} - \frac{1}{8\pi G} \, \int d^{d-1}x \, \int_{z_{P}(\T)}^{\infty} \, \frac{dz}{z^{2}} \, \sqrt{q} \, \theta_{\text{in}}\log\theta_{\text{in}} \nonumber\\ +& \frac{1}{8\pi G} \, \int d^{d-1}x \, \int_{\delta}^{z_{P}(\T)} \, \frac{dz}{z^{2}} \, \sqrt{q} \, \theta_{\text{out}}\log\theta_{\text{out}} \, , \label{act_null}
\end{align}
where $\sqrt{q} = z^{1-d}$ is the determinant of the induced metric of the codimension-$2$ cross-sections of the null hypersurface, and $\theta_{\text{in}}$ ($\theta_{\text{out}}$) is the null expansion of $k_{\text{in}}^{a}$ ($k_{\text{out}}^{a}$). Using Eqs.~\eqref{eq-kin}-\eqref{eq-kout}, we find
\begin{equation}
\theta_{\text{in}} = \, \theta_{\text{out}} = \, (d-1) \, z \, .
\end{equation}
With this result, Eq.~\eqref{act_null} becomes
\begin{align}
\A_{(v>0)}^{\text{null}} =& \, \frac{L_{d-1}}{4\pi G} \,\frac{1}{d-1} \frac{1}{\delta^{d-1}} \, \Big( 1 + (d-1) \log\big[ (d-1)\delta \big] \Big) \nonumber\\-& \frac{L_{d-1}}{4\pi G} \,\frac{1}{d-1} \frac{1}{z_{P}^{d-1}(\T)} \, \Big( 1 + (d-1) \log\big[ (d-1) z_{P}(\T) \big] \Big) \, .
\end{align}

Next we consider the cut-off boundary ($z = \delta$) that we have introduced to regulate the UV divergences. The contribution of this timelike boundary to the gravitational action is given by the Gibbons-Hawking-York (GHY) term. The normal vector to the cut-off boundary is
\begin{equation}
s^{a} = \, \delta \big(\partial_{v}\big)^{a} - \delta \big(\partial_{z}\big)^{a} \, .\label{eq-s}
\end{equation}
The GHY term for the cut-off surface is
\begin{equation}
\A_{(v>0)}^{(z=\delta)} = \, \frac{1}{8\pi G} \,  \int d^{d-1}x \, \int_{\T-2\delta}^{\T} dv \, \sqrt{|\gamma|} \, \gamma^{ab}\nabla_{a}s_{b} \, ,
\end{equation}
where $\gamma^{ab} \equiv g^{ab} - s^{a}s^{b}$ is the inverse induced metric on the cut-off surface. 
Solving this surface integral yields
\begin{equation}
\A_{(v>0)}^{(z=\delta)} = \, \frac{L_{d-1}}{8\pi G} \, \frac{2d}{\delta^{d-1}} \, .
\end{equation}

The last boundary that we have to consider is the spacelike spacetime singularity ($z=\infty$). The contribution of this boundary is also given by GHY term. To compute this term, we first consider the spacelike surface at $z = z_{\infty} \gg z_{h}$ and then take the limit $z_{\infty} \to \infty$. The future-directed vector normal to the $z = z_{\infty}$ surface is 
\begin{equation}
n^{a} = \, -\frac{z_{h}^{d/2}}{z_{\infty}^{d/2-1}} \, \big(\partial_{v}\big)^{a} - \frac{z_{\infty}^{d/2+1}}{z_{h}^{d/2}} \, \big(\partial_{z}\big)^{a} \, .
\end{equation}
The GHY term is then given by
\begin{equation}
\A_{(v>0)}^{(z=\infty)} = \, \lim_{z_{\infty}\to \infty} \, - \frac{1}{8\pi G} \,  \int d^{d-1}x \, \int_{0}^{\T} dv \, \sqrt{|\widetilde{\gamma}|} \, \, \widetilde{\gamma}^{ab}\nabla_{a}n_{b} \, ,\label{eq-ghy-2}
\end{equation}
where $\widetilde{\gamma}^{ab} \equiv g^{ab} + n^{a}n^{b}$ is the inverse induced metric on the $z=z_{\infty}$ surface. The negative sign in Eq.~\eqref{eq-ghy-2} is due to the fact that the shaded region is to the past of the $z=z_{\infty}$ boundary. (See \cite{Lehner:2016vdi} for a summary of the rules to assign the sign to each of the terms in the gravitational action.) Solving the integrals in Eq.~\eqref{eq-ghy-2} yields
\begin{equation}
\A_{(v>0)}^{(z=\infty)} = \, \frac{L_{d-1}}{16\pi G} \, \frac{d}{z_{h}^{d}} \, \T \, .
\end{equation}

Lastly, we consider the contributions of the corner terms to $\A_{(v>0)}$. The corners where null boundaries $v=0$ and $v=\T$ intersect the singularity do not contribute because the volume density of the codimension-$2$ corners falls as $z^{1-d}$. The total contribution to $\A_{(v>0)}$ from the corners is \cite{Lehner:2016vdi}
\begin{equation}
\A_{(v>0)}^{\text{corner}} = \, -\frac{1}{8\pi G} \int_{A} d^{d-1}x \, \sqrt{q} \, a_{A} - \, \frac{1}{8\pi G} \int_{B} d^{d-1}x \, \sqrt{q} \, a_{B} + \, \frac{1}{8\pi G} \int_{P} d^{d-1}x \, \sqrt{q} \, a_{P} \, ,
\end{equation}
where $\sqrt{q} = z^{1-d}$ is the determinant of the induced metric of the codimension-$2$ corners, and 
\begin{align}
a_{A} =& \, \log|k_{\text{in}}\cdot s| \, , \\
a_{B} =& \, \log|k_{\text{out}}\cdot s| \, , \\
a_{P} =& \, \log \left|\frac{ k_{\text{in}}\cdot k_{\text{out}}}{2}\right| \, .
\end{align}
Using Eqs.~\eqref{eq-cor-A}-\eqref{eq-cor-z} and Eq.~\eqref{eq-kin}-\eqref{eq-s}, we get
\begin{equation}
\A_{(v>0)}^{\text{corner}} = \, -\frac{L_{d-1}}{8\pi G} \, \frac{2}{\delta^{d-1}} \, \log\delta \, + \frac{L_{d-1}}{8\pi G} \, \frac{1}{z^{d-1}_{P}(\T)} \Bigg(2 \log z_{P}(\T) - \log f\big(z_{P}(\T)\big) \Bigg) \, .
\end{equation}

The total gravitational action of the shaded region on the left side of Fig.~(\ref{fig-2}) is given by the sum of the contributions from the bulk, surfaces, and corners. For $d=2$, we use the expression of $z_{P}(\T)$ from Eq.~\eqref{eq-z1-2} to get
\begin{equation}
\A_{(v>0)}^{(d=2)} = \, \frac{L_{1}}{4\pi G} \, \frac{1}{\delta} + \, 4\, E_{2} \, \,  z_{h} \, \coth \left( \frac{\T}{2 z_{h}} \right) \, \log \Bigg(  \cosh \left( \frac{\T}{2 z_{h}} \right)\Bigg)    \, , \label{eq-ac-1-d2}
\end{equation}
where we have used Eq.~\eqref{eq-E}. We also compute the time derivative of $\A_{(v>0)}$ for general dimensions. Using the differential equation for $z_{P}(\T)$
\begin{equation}
\frac{d}{d\T} z_{P}(\T)  = \, \frac{1}{2}\, f\big(z_{P}(\T)\big) \, , \label{eq-zp-de}
\end{equation}
we get
\begin{equation}
\frac{d}{d\T}\A_{(v>0)} = \, 2 \, E_{d} \, \Bigg\{ 1 + \left(\frac{z_{h}^{d}}{z_{P}^{d}(\T)} -1 \right)\, \Bigg(  \log\Big(d-1\Big) + \frac{1}{2}\log\Big(f\big(z_{P}(\T)\big)\Big) \Bigg) \Bigg\} \, , \label{eq-adot-1}
\end{equation} 
where we have once again used Eq.~\eqref{eq-E}.

This finishes our discussion of the gravitational action of the shaded region on the left column of Fig.~(\ref{fig-2}). Next, we repeat the analysis of this subsection for the shaded region on the right column of Fig.~(\ref{fig-2}).

\subsection{Calculation of $\A_{(v<0)}$} \label{vac-ads}

We now consider the shaded region on the right side of Fig.~(\ref{fig-2}). This shaded region has three boundaries. One of these is the Poincar\'{e} horizon ($z=\infty$), the other is the infalling null shell ($v=0$), and the third is the null boundary that connects plane $P$ with the past Poincar\'{e} horizon. This third boundary is described by the equation
\begin{equation}
z = \, z_{P}(\T) - \frac{v}{2} \, .\label{eq-z2}
\end{equation}

The bulk contribution to $\A_{(v<0)}$ is given by the Einstein-Hilbert term. This term is proportional to the volume of the shaded region on the right side of Fig.~(\ref{fig-2}). That is,
\begin{equation}
\A_{(v<0)}^{\text{bulk}} = \, \frac{1}{16\pi G} \, \Big( R - 2 \Lambda \Big) \, \V_{(v<0)} \, ,
\end{equation}
where the volume is
\begin{align}
\V_{(v<0)} =& \, \int d^{d-1}x \, \int_{-\infty}^{0} dv \, \int_{z_{P}(\T) - \frac{v}{2}}^{\infty} dz \, \frac{1}{z^{d+1}} \, , \nonumber\\
=& \, \frac{L_{d-1}}{d(d-1)}\,\frac{2}{z_{P}^{d-1}(\T)} \, . \label{eq-vol2}
\end{align}
Using $R-2\Lambda = -2d$, we get
\begin{align}
\A^{\text{bulk}}_{(v<0)} = \, -\frac{L_{d-1}}{4\pi G (d-1)}\, \frac{1}{z_{P}^{d-1}(\T)} \, . \label{eq-bulk-2}
\end{align}

We now consider the boundary contributions to $\A_{(v<0)}$. Note that the Poincar\'{e} horizon ($z=\infty$) does not contribute to $\A_{(v<0)}$ \cite{comp-formation}. 
 For the null boundary at $v=0$, we use the following normalization of the null generator
\begin{equation}
\bar{k}_{\text{in}}^{a} = \, - z^{2} \, \big( \partial_{z}\big)^{a}  \, , \label{eq-kin-2}
\end{equation}
and for the other null boundary, 
we use the following normalization of the null generator
\begin{equation}
\bar{k}_{\text{out}}^{a} = \, 2 \, z^{2} \, \big( \partial_{v}\big)^{a} - z^{2} \, \big( \partial_{z}\big)^{a} \, . \label{eq-kout-2}
\end{equation}
As was the case in Sec.~(\ref{sec-bh}), the surface gravity of these null generators vanish. Therefore, we only have to focus on the the null counter terms \cite{Lehner:2016vdi,Reynolds:2016rvl}. The contributions from the counter terms is \cite{Lehner:2016vdi}
\begin{equation}
\A_{(v<0)}^{\text{null}} = \, \frac{1}{8\pi G} \, \int d^{d-1}x \,  \int_{z_{P}(\T)}^{\infty} \, \frac{dz}{z^{2}} \, \sqrt{q} \, \bar{\theta}_{\text{in}}\log\bar{\theta}_{\text{in}} + \frac{1}{8\pi G} \, \int d^{d-1}x \,  \int_{z_{P}(\T)}^{\infty} \, \frac{dz}{z^{2}} \, \sqrt{q} \, \bar{\theta}_{\text{out}}\log\bar{\theta}_{\text{out}} \, , \label{act_null_two}
\end{equation}
where $\bar{\theta}_{\text{in}}$ and $\bar{\theta}_{\text{out}}$ are null expansion of $\bar{k}_{\text{in}}^{a}$ and $\bar{k}_{\text{out}}^{a}$ respectively. Using $\bar{\theta}_{\text{in}} = \bar{\theta}_{\text{out}} = (d-1)z$, we simplify Eq.~\eqref{act_null_two} as
\begin{equation}
\A_{(v<0)}^{\text{null}} = \, \frac{L_{d-1}}{4\pi G} \,\frac{1}{d-1} \frac{1}{z_{P}^{d-1}(\T)} \, \Big( 1 + (d-1) \log\big[ (d-1) z_{P}(\T) \big] \Big) \, .
\end{equation}

Now let's consider the corner terms. The corners at the intersection of the Poincar\'{e} horizon with the null boundaries do not contribute because the volume density of the codimension-$2$ surfaces falls off as $z^{1-d}$. Therefore, the only corner contribution comes from the corner $P$ and is given by \cite{Lehner:2016vdi}
\begin{equation}
\A_{(v<0)}^{\text{corner}} = \, - \, \frac{1}{8\pi G} \int_{P} d^{d-1}x \, \sqrt{q} \, \bar{a}_{P} \, , 
\end{equation}
where $\sqrt{q}$ is the volume density of the codimension-$2$ corner, and
\begin{align}
\bar{a}_{P} =& \, \log \left|\frac{ \bar{k}_{\text{in}}\cdot \bar{k}_{\text{out}}}{2}\right| \, .
\end{align}
With this, we get
\begin{equation}
\A_{(v<0)}^{\text{corner}} = \, - \frac{L_{d-1}}{4\pi G} \, \frac{1}{z_{P}^{d-1}(\T)} \, \log\big(z_{P}(\T)\big) \, . \label{eq-cor-2}
\end{equation}

After calculating all the contributions to the gravitational action of the shaded region on the right side of Fig.~(\ref{fig-2}), we add them to get
\begin{equation}
\A_{(v<0)} = \, \frac{L_{d-1}}{4\pi G} \, \log\Big(d-1\Big) \, \frac{1}{z_{P}^{d-1}(\T)}  \, . \label{eq-ac-2}
\end{equation}
In the case of $d=2$, the above result vanishes. That is
\begin{equation}
\A_{(v<0)}^{(\text{d=2})} = \, 0 \, . \label{eq-ac-2-d2}
\end{equation}
To find the contribution to the rate of complexification from the shaded region on the right side of Fig.~(\ref{fig-2}), we take the time derivative of Eq.~\eqref{eq-ac-2} 
\begin{equation}
\frac{d}{d\T} \A_{(v<0)} = \, - 2 \, E_{d} \, \left(\frac{z_{h}^{d}}{z_{P}^{d}(\T)} -1 \right)\,  \log\Big(d-1\Big) \, , \label{eq-adot-2}
\end{equation}
where we have once again used Eq.~\eqref{eq-zp-de}.

In the next subsection, we combine the results of this and the previous subsections to find the total action of the WdW patch and hence, the complexity of the boundary state.

\subsection{Calculation of the total action} \label{total-action}

The total gravitational action of the WdW patch is given by Eq.~\eqref{eq-ac-total}. In the case of $d=2$, we add Eq.~\eqref{eq-ac-1-d2} and Eq.~\eqref{eq-ac-2-d2} to get
\begin{equation}
\A^{(\text{d=2})} = \, \frac{L_{1}}{4\pi G} \, \frac{1}{\delta} + \, 4\, E_{2} \, \,  z_{h} \, \coth \Big( \frac{\T}{2 z_{h}} \Big) \, \log \Bigg(  \cosh \left( \frac{\T}{2 z_{h}} \right)\Bigg) \, .
\end{equation}
The CA conjecture, Eq.~\eqref{CA-conj}, then implies that the complexity of the perturbed state at time $\T$ is given by
\begin{equation}
\C^{(\text{d=2})} (\T) = \, \frac{L_{1}}{4\pi^{2} G} \, \frac{1}{\delta} + \, \frac{4}{\pi}\, E_{2} \, \,  z_{h} \, \coth \Big( \frac{\T}{2 z_{h}} \Big) \, \log \Bigg(  \cosh \left( \frac{\T}{2 z_{h}} \right)\Bigg) \, . \label{comp-d2}
\end{equation}
The analytical expression for the time-dependence of the complexity in $d=2$ is one of the main results of this paper. For $d\ge 3$, we do not have the analytical result for the time-dependence of the complexity. The reason for this is that we do not have the closed form solution of Eq.~\eqref{def-z1}. In spite of that, we can still study the time evolution of the complexity in $d\ge 3$ by studying its rate of growth. We add Eq.~\eqref{eq-adot-1} and Eq.~\eqref{eq-adot-2} and use CA conjecture to get
\begin{equation}
\frac{d}{d\T}\, \C(\T) = \, \frac{2}{\pi} \, E_{d} \, \Bigg\{ 1 +\frac{1}{2} \left(\frac{z_{h}^{d}}{z_{P}^{d}(\T)} -1 \right)\, \log\Big(f\big(z_{P}(\T)\big)\Big) \Bigg\} \, ,\label{comp-der}
\end{equation}
We use this expression to study the bound on the rate of complexification. As is evident from Fig.~(\ref{fig-1}) and from the left side of Fig.~(\ref{fig-2}), the plane $P$ never crosses the event horizon (\textit{i.e.} $z_{P}(\T) \le z_{h}$). This means that the second term inside the parenthesis on the right side of Eq.~\eqref{comp-der} can never be positive. This gives us the following bound
\begin{equation}
\frac{d}{d\T}\, \C (\T) \le \, \frac{2}{\pi} \, E_{d} \, , \label{comp-rate-bound}
\end{equation}
which is in agreement with the bound of Eq.~\eqref{eq-Lloyd-bound}. Furthermore, note that the plane $P$ approaches the event horizon (\textit{i.e. }$z_{p}(\T) \to z_{h}$) at late times ($\T \gg z_{h}$). This means that the above inequality eventually saturates. This analytical consistency check of the CA conjecture and the conjectured bound on the rate of complexification is another main result of this paper.

\begin{figure}[h]
	\begin{tikzpicture}
	\node at (0,0) {\includegraphics[scale=0.6]{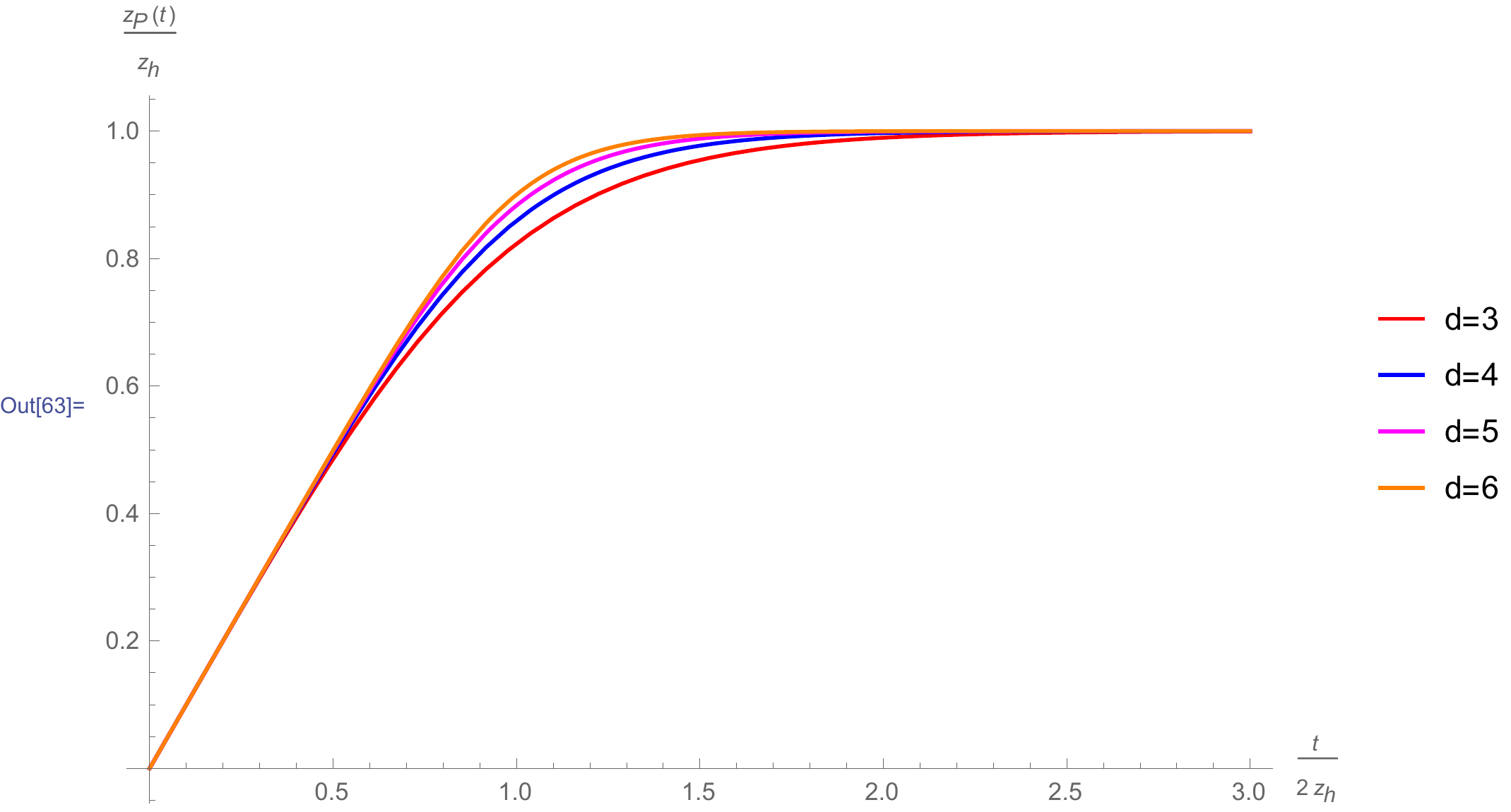}};
	\fill [white](-5.250,3.24) circle (.4cm);
	\node at (-5.25,3.1) {$z_{P}(\T)/z_{h}$};
	\fill [white](5.0,-3.20) circle (.4cm);
	\node at (4.9,-3.2) {$\frac{\T}{2z_{h}}$};
	\fill [white](-6.150,0.15) circle (.4cm);
	\end{tikzpicture}
	\caption{The plots showing the position of the plane $P$ as a function of time for $d = \{3,4,5,6\}$. It is evident from these plots that $z_{P}(\T) \to z_{h}$ when $\T \sim z_{h}$.}
	\label{fig-3}
\end{figure}
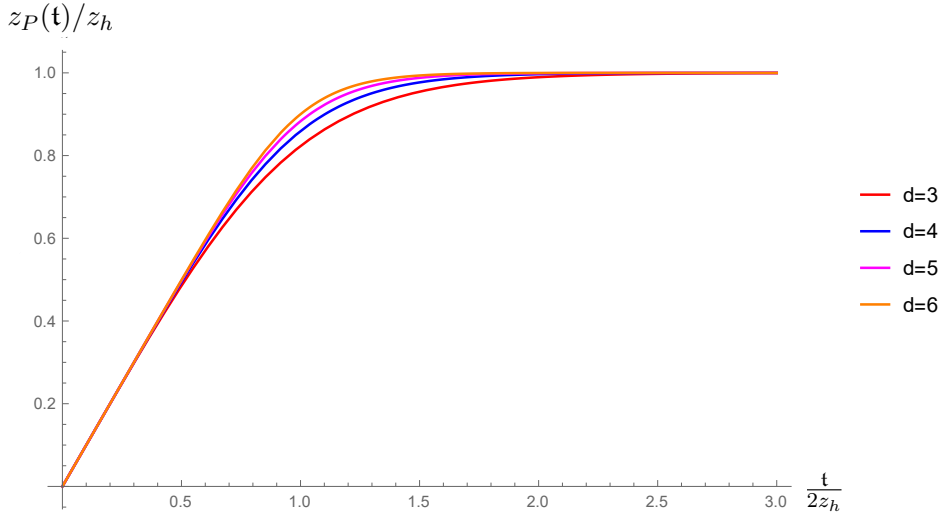

Another interesting question that one may ask is how long does it take for the rate of growth of complexity to approach its maximum possible value. To answer this question, we need to find how long does it take for the plane $P$ to reach the event horizon. Recall that the function $z_{P}(\T)$ satisfies the differential equation, Eq.~\eqref{eq-zp-de}, with the boundary condition $z_{P}(\T=0) = 0$ (see Eq.~\eqref{def-z1}). We numerically solve this differential equation for $d= \{ 3, 4, 5, 6\}$. The plots of $z_{P}(\T)$ as a function of time are shown in Fig.~(\ref{fig-3}). It is clear from these plots that the plane $P$ approaches the event horizon when $\T \sim z_{h}$. This time scale is same as the inverse temperature of the CFT state (see Eq.~\eqref{eq-E}) or the local equilibrium scale. This result, together with Eq.~\eqref{comp-der}, implies that the rate of complexification attains its maximum value soon after the system has achieved local equilibrium. 

\section{Discussion} \label{disc}

In this paper, we excited the vacuum state of a $d$-dimensional CFT by a global quench and studied the growth of the complexity of the resultant time-dependent state. We quenched the system by injecting homogeneous and isotropic energy globally in the system. In the gravity side of AdS-CFT correspondence, this quench is described as an infalling null shell which collapses to form a black hole \cite{gq-3,gq-4,gq-6,gq-7,gq-8,gq-10,gq-11,gq-13,gq-14}.  We used the CA conjecture, Eq.~\eqref{CA-conj}, to study the growth of the complexity of the boundary state.

For $d=2$, we derived an expression of the time-dependence of the complexity. This expression is given in Eq.~\eqref{comp-d2}. For $d \ge 2$, we derived the rate of growth of complexity in terms of single time-dependent function, $z_{P}(\T)$. This is given in Eq.~\eqref{comp-der}. The function $z_{P}(\T)$ is the position of the codimension-$2$ plane where the WdW patch corresponding to boundary state at time $\T$ intersects the infalling null shell. One can deduce from Fig.~(\ref{fig-1}) and Fig.~(\ref{fig-2}) that the plane $P$ never crosses the event horizon. That is, $z_{P}(\T) \le z_{h}$. With this inequality, our general expression for the rate of complexification reduces to the conjectured bound of Eq.~\eqref{eq-Lloyd-bound}. Furthermore, we found the time-dependence of $z_{P}(\T)$ by numerically solving Eq.~\eqref{eq-zp-de}. The plots for $z_{P}(\T)$ for $d = \{3, 4, 5, 6\}$ are shown in Fig.~(\ref{fig-3}). This numerical analysis allowed us to deduce that the quenched state saturates the bound on the rate of complexification soon after the system has achieved local equilibrium.

We now discuss some possible directions in which the present work can be extended. 
\begin{enumerate}
	\item Consider a more realistic protocol of quantum quenches, where one perturbs the CFT by introducing a relevant operator with a time dependent coupling. That is, 
	\begin{equation}
	H_{\text{CFT}} \to \, H(\T) = \, H_{\text{CFT}} + \lambda(\T) \, \mathcal{O} \, ,
	\end{equation}
	where $\lambda(\T) = 0$ for $\T \le 0$ and $\mathcal{O}$ is a relevant operator. The holographic description of this quench involves an introduction of the scalar field in the bulk with the boundary condition governed by the time-dependent coupling, $\lambda(\T)$. The computation of the on-shell action of the WdW patch will require the solutions of the coupled Einstein-Klein-Gordon equations. In general, this is a difficult calculation. However, one can try solving the bulk equations near the boundary using Fefferman-Graham ansatz, and use it to extract the time dependence, if any, of the divergences associated with the gravitational action of the WdW patch. It will be interesting to investigate how the bound on the rate of complexification works if the complexity has time-dependent divergences. [Note: See \cite{Moosa:2017yiz} for a recent calculation of these time-dependent UV divergences.]
	
	\item The complexity equals action was generalized in \cite{comp-subregions} for reduced states of some subsystem of the CFT. This proposal relates the complexity of the reduced state of some boundary subregion $A$ with the on-shell action of the bulk region defined as the intersection of the entanglement wedge of $A$ and the WdW patch of a boundary slice that includes region $A$. One may consider repeating the analysis of this paper for the time-dependence of the complexity of some fixed boundary region. However, it is known \cite{Albash:2010mv,gq-9,gq-10} that there can be discontinuities in the position of the HRT surface \cite{Hubeny:2007xt}, and consequently in the entanglement wedge, when the time $\T$ is of the order of the size of the boundary subregion. This is due to the presence of more than one boundary anchored extremal surfaces. This jump in the HRT surface of the subregion can result in the discontinuity of the complexity of the reduced state of the subregion. It will be interesting to investigate if this really happens, and if it does, then how it is compatible with the bound on the rate of complexification.
\end{enumerate}


\vskip .3cm
\indent {\bf Acknowledgments} 
It is a pleasure to thank Ning Bao, Raphael Bousso, Adam Brown, and Pratik Rath for helpful discussions, and to Saad Shaukat for useful
feedback on a draft of this manuscript. This work was supported in part by the Berkeley Center for Theoretical Physics, by the National Science Foundation (award numbers 1521446 and 1316783), by FQXi, and by the US Department of Energy under Contract DE-AC02-05CH11231.

\bibliographystyle{utcaps}
\bibliography{all}
\end{document}